\documentclass[journal]{IEEEtran}
\IEEEoverridecommandlockouts
\usepackage[pdftex]{graphicx}
\usepackage{amsmath}
\usepackage{algorithmic}
\usepackage{graphicx}
\usepackage{textcomp}
\usepackage{xcolor}
\begin{document}
\bstctlcite{IEEEexample:BSTcontrol}
\title{Telecom Foundation Models:\\ Applications, Challenges, and Future Trends}

\author{\IEEEauthorblockN{Tahar Zanouda\,, 
Meysam Masoudi\,, 
Fitsum Gaim Gebre, Mischa Dohler }\\
\IEEEauthorblockA{Ericsson}\\
\texttt{\{tahar.zanouda, meysam.masoudi, fitsum.gaim.gebre, mischa.dohler\}@ericsson.com}}

% Target Journal: IEEE Communications Standards Magazine 
% https://www.comsoc.org/publications/magazines/ieee-communications-standards-magazine/cfp/ai-wireless

\maketitle

\begin{abstract}
% introduce a new name - Telco-FMs
%  Telecom Foundation Models (TFMs)

Telecom networks are becoming increasingly complex, with diversified deployment scenarios, multi-standards, and multi-vendor support. The intricate nature of the telecom network ecosystem presents challenges to effectively manage, operate, and optimize  networks. To address these hurdles, Artificial Intelligence (AI) has been widely adopted to solve different tasks in telecom networks. However, these conventional AI models are often designed for specific tasks, rely on extensive and costly-to-collect labeled data that require specialized telecom expertise for development and maintenance. The AI models  usually fail to generalize and support diverse  deployment scenarios and applications. In contrast, Foundation Models (FMs) show effective generalization capabilities in various domains in language, vision, and decision-making tasks. FMs can be trained on multiple data modalities generated from the telecom ecosystem and leverage specialized domain knowledge. Moreover, FMs can be fine-tuned to solve numerous specialized tasks with minimal task-specific labeled data and, in some instances, are able to leverage context to solve previously unseen problems. 
At the dawn of 6G, this paper investigates the potential opportunities of using FMs to shape the future of telecom technologies and standards. In particular, the paper outlines a conceptual process for developing Telecom FMs (TFMs) and discusses emerging opportunities for orchestrating specialized TFMs for network configuration, operation, and maintenance. Finally, the paper discusses the limitations and challenges of developing and deploying TFMs.

%Such a transformation mandates a 

\end{abstract}

%\bstctlcite{IEEEexample:BSTcontrol}

\section{Introduction} 

5G-and-beyond networks offer immense opportunities across various industries, enabling limitless connectivity and numerous emerging use cases. However, the increasing complexity of mobile networks hinders future developments, with deployment scenarios ranging from centralized to virtualized and physical implementations, multi-vendor support, and open ecosystem. To address these impediments, AI has been adopted in telecom to enhance automation and reduce manual tasks in network operations, administration, and maintenance, leading to the inception of \textit{TelcoAI - AI for Telecom}.

As we move towards 6G \cite{TwelveChallenges6G}, AI/ML is expected to be a foundational technology, driving AI-centric networks and building on the extensive use of AI in current networks throughout their design, deployment, and operational stages. The growing importance of AI in telecom is portrayed by the
concept of \textit{"AI-native Telecom"}, where AI is at the core of network functionalities. Recently, AI has been applied to various tasks, including ML-based modulation schemes, radio resource management (e.g., power control, scheduling, link adaptation), UE localization, mobility optimization, network slicing, network configuration, and wireless security. Graph Neural Networks (GNNs) incorporate spatio-topological dynamics in the network. Reinforcement Learning (RL) leverages feedback through exploration-exploitation learning techniques using online and offline RL architectures. Transfer Learning addresses data scarcity by transferring knowledge from related tasks. Similarly, label-efficient learning and few-shot learning mitigate data scarcity and labeling challenges.

Given data sensitivity and limited  resources, distributed ML has become integral to \textit{TelcoAI}. Federated Learning allows remote clients to collaboratively train models in a privacy-preserving manner, using isolated data and predefined aggregation strategies to update local models. Split Learning decouples models into parts trained on local data, with outputs fused into a central model on the server side.
These advancements highlight the potential of AI and foundation models in telecom, paving the way for more sophisticated models to harness the full potential of 6G and beyond \cite{DistributedFM}. %Key factors propelling \textit{TelcoAI}'s evolution include increased real data availability, enhanced processing power, and improved intelligent RAN systems. 

However, several challenges arise when applying AI/ML in real-world telecom applications, such as limited ability to generalize, inability to capture network complexities, necessity to train models on data generated by network simulators, among others. To this end, Foundational Models (FMs) can be enablers for autonomous telecom networks \cite{bommasani2021opportunities}, eliminating the necessity of task-based AI models while addressing the current limitations.

%\begin{itemize}
%   \item \textit{Limited Generalization:} Traditional deep learning models in telecom, trained on small datasets tailored for specific tasks, struggle to adapt across diverse deployment scenarios, resulting in marginal improvements.

%    \item \textit{Telecom Network Diversity:} Telecom networks are heterogeneous, supporting multiple Radio Access Technologies (RATs), diverse deployment scenarios, open and multi-vendor technologies, and diverse service requirements, making capturing their complexity and variability difficult.
%    \item \textit{Re-usability of \textit{TelcoAI} Solutions:} Developing \textit{TelcoAI} models for different tasks across the value chain often involves training models on similar data inputs for various downstream tasks (e.g., software assessment during testing and field operations). Managing multiple ML solutions across large distributed telecom vendors escalates operational expenses. Adopting a holistic view of downstream tasks and their data inputs can streamline the MLOps stack.
%    \item \textit{Sim2Real Gap:} Due to a lack of data, \textit{TelcoAI} solutions often resort to training models on network simulators that do not capture the complexities of real networks.
%\end{itemize}
%While AI methods demonstrate effectiveness, evolving methodologies are necessary to address diverse telecom network challenges. With recent advancements in Foundation Models (FMs) \cite{bommasani2021opportunities}, 

%to generalize and support diverse telecom deployment scenarios. 

FMs are large models trained on massive and heterogeneous datasets to solve a multitude of downstream tasks \cite{intentLLM, GenerativeAITelecom}. They are versatile and can be deployed for numerous use cases in telecom and other domains. FMs can be categorized based on data modality and learning process as follows: 
\begin{itemize}
    \item \textit{Large Language Models (LLMs)} are large models trained on a large corpus of textual data. Models such as LLaMA, PaLM, GPT-4, Falcon, and Mistral, among others, showed impressive results in the range of tasks. LLMs can also be fine-tuned with domain-specific data to align the performance of the LLMs to a specific domain such as Telecom \cite{GenerativeAITelecom}. LLMs have been used for different tasks in SW engineering, such as building chatbot assistants, code generation, and maintenance ticket resolution \cite{fan2023large}.

    \item \textit{Large Vision Models} are large models designed to solve several computer vision tasks. Models such as ViT, Meta's DINO, Meta's SAM, and Florence showed promising results. Other models such as Prithvi were trained on geospatial and satellite imagery data.

    \item \textit{Time-Series Models} are large models designed to train a unified model for time-series. Models such as LLM-Time, Lag-LLAMA, LLM4TS, and LLMTime are early efforts in this area.

    \item \textit{Multimodal FMs} are large models designed to deal with multimodal data that refer to the ability of a model to accept different input modalities, e.g., images, texts, or audio signals. OpenAI’s CLIP and DALL-E, BLIP, FILIP, PaLM, LLaVA and GPT-4o are examples of such models.

    \item \textit{RL FMs} are large models that combine FMs and sequential decision-making to tackle complex real-world problems with better generalization. Recently, attempts have been made to build upon decision transformers or introduce tailored architectures, such as DeepMinds's Gato to achieve multi-modal, multi-task, and multi-embodiment objectives such as playing Atari, captioning images, chatting, etc. 
\end{itemize}

This paper delves into the development of Telecom FMs (TFMs) and briefly summarizes the data types and modalities in the telecom ecosystem. The article describes FM use cases in telecom and sheds light on the challenges and risks of deploying TFMs.

\section{AI/ML Standards \& Alliances for RAN}

Integrating AI into 5G and beyond networks requires developing an aligned view through standardization bodies to develop AI technology standards.  
Multiple alliances have been established by partnering with technology industry leaders and academic institutions to enhance RAN performance and capabilities using AI \cite{chaccour2024telecom}. These alliances aim to capitalize on recent generative AI capabilities and accelerate the adoption of \textit{GenAI} in the telecom sector to find new growth opportunities. 

\begin{itemize}
    \item \textit{AI RAN Alliance} a collaborative initiative to develop AI-driven solutions to achieve an AI-native RAN.

    \item \textit{Global Telco AI Alliance (GTAA)} a collaboration between key telecom operators to advance AI use cases reshaping the telecom landscape.
        
    \item \textit{Alliance for Telecommunications Industry Solutions (ATIS)} ATIS-TOPS Council established a working group to evaluate GenAI/ML use cases for future networks, examining how it contributes to improving the efficiency across Telecom sectors.

\end{itemize}
 
Standards defining organizations (SDOs) aim to standardize the application of AI in telecom, paving the way for more efficient, secure, and intelligent network operations.
Such initiatives are as follows:
\begin{itemize}
    
    \item \textit{O-RAN} is investigating cross-domain AI and Generative AI use cases in Open RAN architecture. O-RAN specifications are being adopted by the European Telecommunication Standardization Institute (ETSI) and published among ETSI standards.

    \item \textit{ETSI} established a new AI Agent Core Network working group to discuss AI for next-generation telecom technologies. Key focus areas include Orchestration and management of network operations, network knowledge management, intelligent customer service applications,  LLMs Lifecycle Management, securities, etc. 

    \item \textit{3GPP}  is now integrating AI into RAN. For instance, TS 28.105 (Rel. 17) and TR 28.908
(Rel. 18) outlines AI/ML integration in 5G systems. Ongoing discussions focus on leveraging AI for enhancements in CSI, beam management, and positioning, highlighting 3GPP’s dedication to the responsible integration of AI in wireless communications.

    \item \textit{International Telecommunication Union (ITU} with two groups (1) ITU-T to set global ICT standards, and (2) ITU-R to advance radio technologies with AI integration. ITU-T: Develops global ICT standards focusing on internet protocols, IoT, and next-generation networks (Y-series). Supplement 72 to the Y.3000-series details AI integration in ICT, featuring key standards like Y.3115 for AI-enabled cross-domain network requirements and Y.3325 for an AI-based management framework. ITU-R enhances radio technologies through AI integration, as outlined in Document M.2242. This document emphasizes Cognitive Radio Systems (CRS) with capabilities such as environment awareness, autonomous parameter adjustment, and adaptive learning.
    
\end{itemize}

Subsequently, we highlight opportunities and challenges for these AI-centric SDOs and Alliances to adopt TFMs.

\section{Telecom Data Ecosystem}

\begin{figure*}[ht]
    \centering
    \includegraphics[width = 0.9\linewidth]{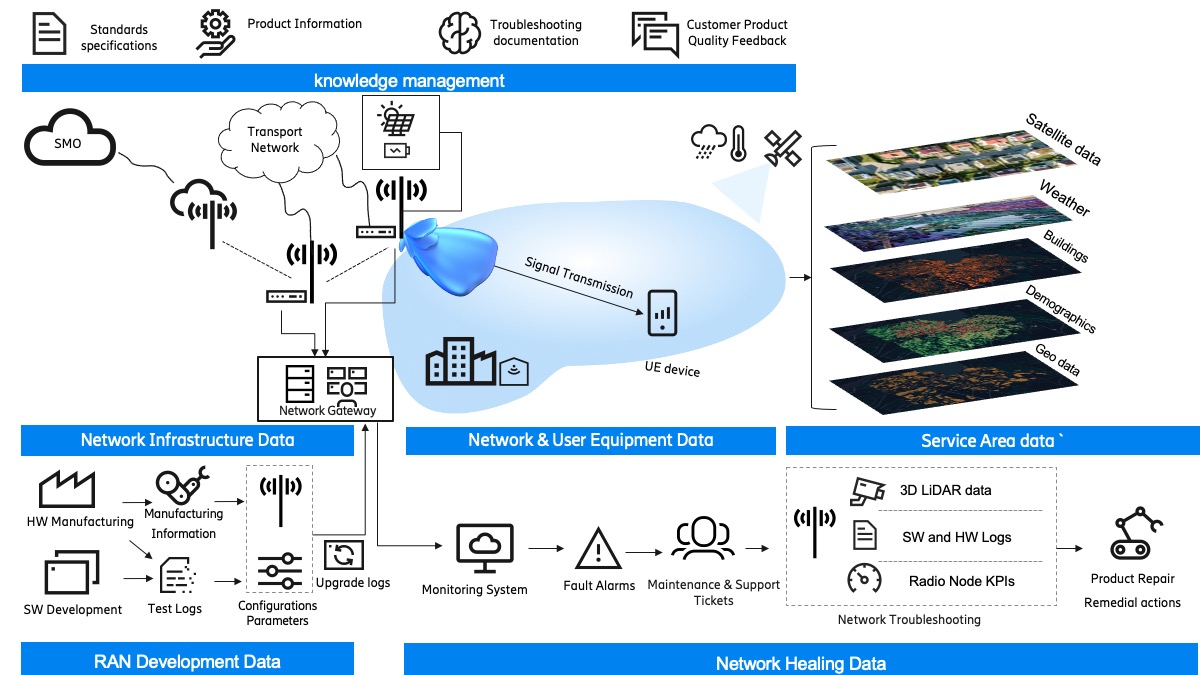}
    \caption{The Overview of Telecom Network and Ecosystem Data.}
    \label{fig:data_ecosystem}
\end{figure*}

Telecom networks generate multimodal data from spatially distributed radio nodes, capturing temporal dynamics originating from different software (SW) and hardware (HW) modules and storing them in various file formats. Figure \ref{fig:data_ecosystem} depicts a simplified overview of the telecom data ecosystem that spans phases of HW manufacturing, SW development, SW/HW testing, product knowledge management, network deployment, and network troubleshooting and healing. 
Telecom networks consist of a set of heterogeneous interconnected radio nodes that are distributed in a geographical region to provide connectivity services. Radio nodes encompass SW and HW components, each providing different functionality. The HW components are configured based on configuration files, and SW components cover the implementations of multiple technology standards. System configuration for both SW and HW components are captured in \textit{radio node configuration data} \cite{lin2023artificial}, which encompass information about telecom networks configuration, including \text{Configuration Management (CM) parameters} \cite{lin2023artificial}, SW features, licenses, and installation setup.

The telecom infrastructure consists of a wide variety of assets, e.g., power equipment, data centers, inventory warehouses, etc, captured in {telecom infrastructure inventory and configuration} data along with a diary of {maintenance activities} such as cleaning filters, inspecting backup batteries, etc. Operators monitor the network continuously to assess the behavior of the network using {Performance Management (PM) data} \cite{lin2023artificial} to gauge network performance. PM data is captured regularly across radio nodes in different SW and HW components. 

Operators benchmark the behavior of the network and its SW using {Key Performance Indicators (KPIs)}, which are standardized by \textit{3GPP}. KPIs are a set of formulas to calculate performance indicators using PM and CM data. Different categories of \textit{KPI}s exist, such as accessibility, mobility, integrity, utilization, and energy performance. This \textit{KPI}s behavior can vary depending on service area characteristics. SW components use such \textit{KPI}s along with other mechanisms to generate alerts for any aberrant behavior in the network. The format of the alerts is standardized by \textit{3GPP} and known as {fault management data} \cite{lin2023artificial}.

Operators monitor network traffic usage and activities through different data sources (e.g., service usage, subscription, call records, etc). The data that characterizes the interaction between users and the network can be captured in several ways. For instance, {minimization of drive test} measurements, standardized by 3GPP \cite{lin2023artificial}, consist of user information, field measurements, radio measurements, and location information.

%Telecom networks provide coverage to end-users. In each service area, User equipment (UE) communicates via RAN. The data that characterizes the interaction between UE and network can be captured in several ways; \textit{Minimization of Drive Test (MDT)} Measurements \cite{lin2023artificial}, standardized by \textit{3GPP} \cite{lin2023artificial}, consists of UE information, field measurements, radio measurements, and location information.

Developing SW-intensive HW products in the telecom typically involves HW manufacturing, SW development, HW/SW integration, testing, and field trials. Network products implement standardization guidelines that outline globally aligned solutions, functional frameworks, use cases, and test specifications captured in {standardization text documents} \cite{lin2023artificial}. RAN HW is typically tested with simulations, and when the HW design is proven to be functional, the units are produced, calibrated, and tested in factories. %\textit{Manufacturing information and configuration} is also to ensure the reliability of HW products. 
The process of HW testing generates {HW test log files} that capture the output of such activities in the factory. During these activities, {manufacturing information} data is stored to ensure the reliability of HW products across the different stages. %{\color{red} The process of manufacturing HW equipment also requires a resilient and sustainable global supply chain to track the interconnected journey of delivering products from factories to networks; all activities generate data that are tracked in \textit{Product Supply Chain}.}

RAN SW comprises several large components referred to as {SW modules}. Each module is developed and maintained by one or several cross-functional teams. 
{SW modules} capture logs, events, and internal counters to enable instant feedback and alerts for any aberrant behavior in the system. {Log files} are a text-based function-related history of events that describe software state during its execution. Each line of log files indicates a different event and may hold various types of information such as function name, timestamp, and log message. Other types of logs, such as SW logs, HW logs, and traces, are typically used for debugging and capturing low-level events. A trace can span multiple functions and be tracked using unique references. 
{SW internal counters}, on the other hand, are time-dependent data points to monitor SW quality. Unlike standardized data,  no common formatting schemes exist for different vendors to convey similar information for SW logs and counters.
When a SW component is deployed in the cloud, cloud resource utilization data, referred to as {telemetry data}, are stored for elastic and dynamic resource management.

Before rolling out any SW release, each SW typically undergoes rigorous testing and review in a {simulator}. The process involves creating many {test cases} managed in {test case specification documents}, each crafted to validate dedicated system features.
Testing SW produces {SW test log files}.

After rolling out SW/HW products in the field, ensuring the product's reliability is crucial. The reporting, analyzing, and resolving HW and SW faults are essential to providing stable, high-quality products. When a radio node experiences a problem/incident in an operational network, the operator raises a {troubleshooting report} or {maintenance request} as a textual document. The trouble report contains problem observations, system logs, and an answer section for resolved reports \cite{bosch2022trdi}. 

The diversity of data generated from Telecom ecosystem presents several challenges, including but not limited to data multimodality with large diversity in size, granularity, etc; quickly outdated data quality; data scarcity for rare but impactful events; dynamics of data feedback loops; etc.

%\begin{itemize}
%    \item \textit{Data Multimodality}: Telecom networks, with multiple deployment options, are inherently distributed and diverse, producing multimodal data such as images, software time-series data, or textual logs. This leads to a large diversity of data in terms of size, granularity, format, and type. 
%    \item \textit{Data Quality}: Telecom's diverse ecosystem, domain experts' assistance requirements, and quickly outdated data limit the data quality. 
%    \item \textit{Data Scarcity}: Given the hardware-constrained environment of RAN and the diversity of RAN deployment scenarios, RAN can experience rare events and data loss, contributing to data scarcity.  
%    \item \textit{Data Feedback Loop}: Telecom networks use closed-loop operations in environments that require dynamic network configurations and orchestration. 
%    \item \textit{Data Volume}: Telecom networks generate large amounts of difficult data to handle, process, and store. 
%    \item \textit{Data Privacy}: Telecom networks generate data that is considered sensitive, which presents a challenge to use, store, and process data. 
%\end{itemize}
% \newpage

\section{Telecom Foundation Models } 

\subsection{Developing Telecom Foundation Model}

\begin{figure}[t]
    \centering
    \includegraphics[width = 0.99\linewidth]{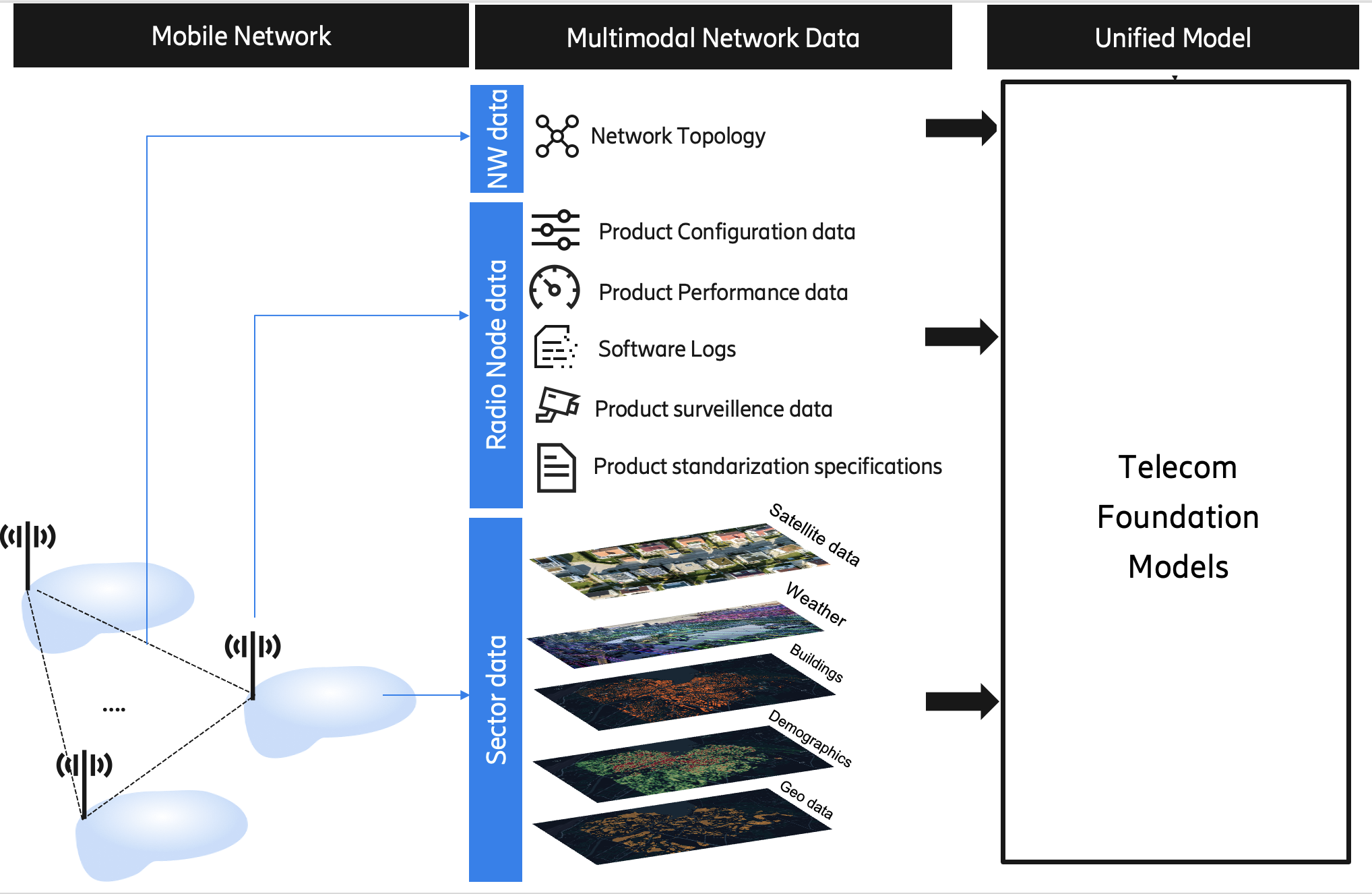}
    \caption{Conceptual Telecom Foundation Model Architecture.}
    \label{fig:tfms}
\end{figure}

Telecom Foundation Models (TFMs) are domain-specific FMs trained on extensive data, spanning multiple sub-domains and modalities, and tailored for telecom applications. Here, we present the necessary components and pipelines for developing a TFM in telecom, which starts by assembling a dataset using multiple and multi-modal data sources. 

Telecom data comes in different temporal, visual, and spatial modalities to build a holistic overview of how telecom networks function jointly. The next step is data fusion, which refers to integrating multiple data sources from various data sets to generate more accurate, comprehensive, and useful information. Telecom data may have multiple time granularity, asynchronous reporting times, and multi-modality.

The training phase of the model requires designing an architecture capable of capturing a comprehensive representation of telecom networks. This architecture conceptualizes telecom data as \textit{context-aware multimodal heterogeneous graphs}. The model incorporates three key components:

\begin{itemize}
\item \textbf{Radio Node Component}: it is conceptualized as a multi-level graph that characterizes software execution, building upon previous work \cite{TroubleshootingGNN, ConfigurationRecGNN}, using (i) software performance time-series data, (ii) textual log messages, and (iii) configuration parameters. Each radio node is further characterized by manufacturing information to understand how sensors and products are jointly affected. In Fig. \ref{fig:tfms}, we show different data types that can be extracted and used in training the model from each radio node.

\item \textbf{Network Component}: it is conceptualized as a multi-modal graph that characterizes interactions between different RAN nodes, building upon previous efforts \cite{ConfigurationRecGNN, TroubleshootingGNN}, using (i) RAN nodes relationships which indicate the relation between each pair of network nodes. The relations may be any one or a combination of a (1) mobility relation, and/or a (2) transport relation between pair of network nodes, and (ii) service area data. This data captures the dynamics of the surrounding area, including geographical (e.g., satellite imagery), socio-demographic (e.g., population density), and economic factors in coverage areas. In Fig. \ref{fig:tfms}, we use the network topology to extract the adjacency matrix that is crucial to train GNN-based architectures.

\item \textbf{Network Development Journey Component}: it is conceptualized as a time-dependent set that can be collected through various network stages: (i) product development (e.g., product guidelines, test logs), (ii) network optimization (e.g., expert rules), (iii) network evolution (e.g., upgrade actions), and (iv) network healing (e.g., maintenance tickets). Such data is typically used when fine-tuning models for specialized tasks, building upon previous efforts \cite{bosch2022trdi, ConfigurationRecGNN, TestCodeGenAI}.
\end{itemize}

A global model is initially trained on general telecom knowledge but is not fine-tuned for specific applications. In the downstream fine-tuning phase, this global model is leveraged in various telecom applications such as network optimization. During downstream tasks, transfer learning or fine-tuning becomes crucial to adapt the model to specific tasks while inheriting the general telecom knowledge from the global model.

\subsection{Specialized Telecom Foundation Models} %: Applications \& Emerging Opportunities
The general TFM, trained with telecom knowledge, can be tailored to different domains. The TFM architecture is illustrated in Figure \ref{fig:tfms}.

Telecom applications are diverse, presenting unique limitations, tasks, and complexities. However, they all rely on core telecom domain knowledge. As a result, a general TFM can be adapted for specific tasks through different techniques. 

To build a specialized model, one can consider the data modality and the use case's nature. These techniques can be broadly grouped into two classes:
\begin{itemize}
    \item Domain Adaptation of Models, which involves training models.
    \begin{itemize}
        \item Pre-training new models from scratch.
        \item Continued pre-training of existing models.
        \item Domain-specific fine-tuning/Instruction tuning of existing models, which involves adjusting the TFM's parameters to generate and optimize outputs for particular applications. 
        \end{itemize}
    
    \item In-context Learning and Knowledge Augmentation of existing models, which does not involve training models. The approaches are applicable to both general and domain-specific models.
    \begin{itemize}
        \item Prompt engineering, which uses techniques such as Zero-Shot, One-Shot, and Few-Shot Learning. 
        \item Retrieval Augmented Generation (RAG) and Graph Retrieval Augmented Generation (Graph-RAG).
    \end{itemize}
    
 \end{itemize}

% Moreover, there are techniques that improve the output of FMs, such as 
%Reinforcement learning from human feedback (RLHF), which helps improve the model's accuracy by incorporating feedback on model behavior.

These techniques are mostly used in tandem with the FMs. Other techniques help reduce the cost of model fine-tuning and management. Several techniques have emerged recently. For instance, Low-rank adaptation (LoRA) significantly reduces large models' trainable parameters and GPU memory requirements. LoRA enhances speed, reduces compute resource demands and cost, and improves memory efficiency by allowing parameters to be cached in memory instead of relying on slower disk reads.

%%%%%%

\begin{figure*}[ht]
\centering
    \includegraphics[width =\linewidth]{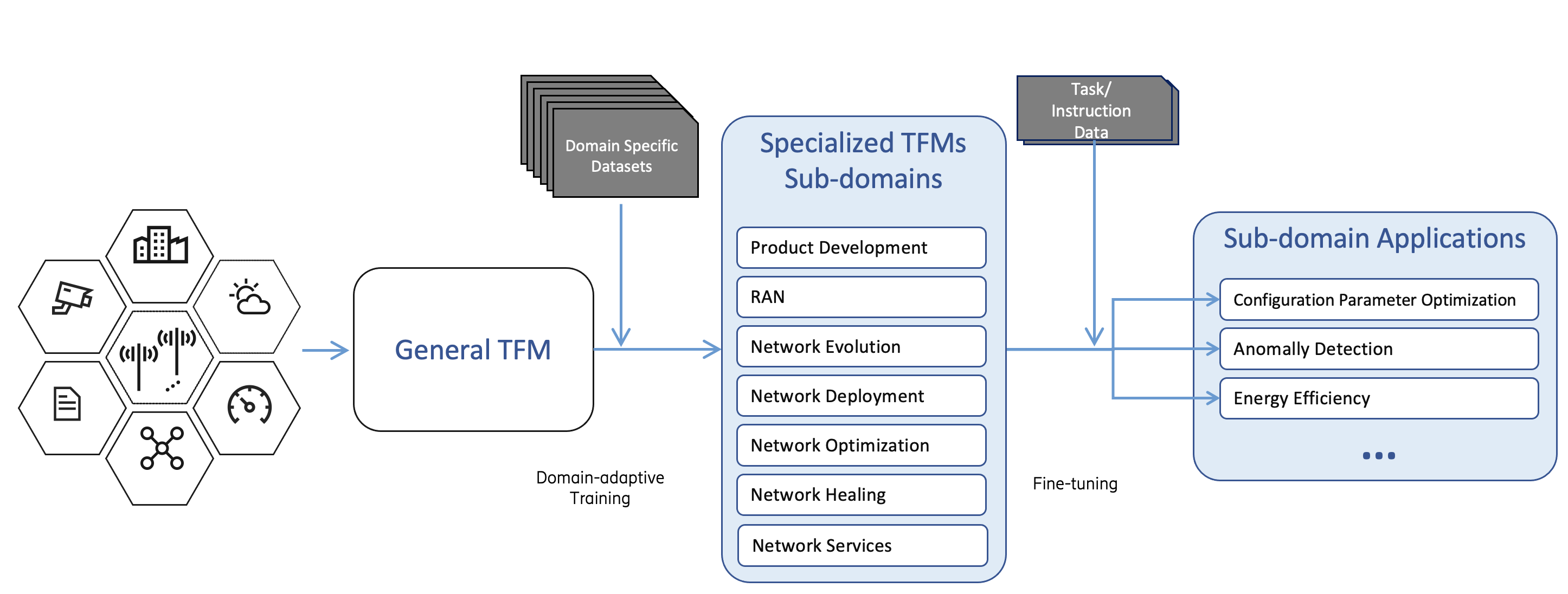}
    %fine_tuning_v2
    \caption{Building specialized Telecom foundation models for each area}
    \label{fig:specialized_tfm}
\end{figure*}

%Fine-tuning involves adjusting the TFM's parameters to generate and optimize outputs for particular applications. Depending on the use case and the chosen foundation model, there are two main approaches to parameter-efficient fine-tuning of TFMs:
% \begin{itemize}
%     \item Instruction-based fine-tuning using prompts: This method uses labeled examples to enhance the performance of a pre-trained foundation model on a specific task. These labeled examples are formatted as prompts or response pairs and phrased as instructions, guiding the model to perform the desired task more effectively.
%     \item Domain-specific data fine-tuning: If prompt engineering does not provide sufficient customization, domain adaptation fine-tuning can be employed. This method involves training the model with domain-specific languages, such as industry jargon, technical terms, or other specialized data, to improve its performance on tasks within that domain. For instance, Low-rank adaptation (LoRA) significantly reduces large models' trainable parameters and GPU memory requirements. LoRA enhances speed, reduces compute resource demands and cost, and improves memory efficiency by allowing parameters to be cached in memory instead of relying on slower disk reads.
%     \item Graph Retrieval-Augmented Generation (RAG).
% \end{itemize}
Telecom has many applications, and each specialized TFM is tailored to specific domains that encompass similar tasks and require similar inputs. However, fine-tuning a TFM is a resource-intensive and costly process. Therefore, specialized TFMs are designed to cover categorized tasks efficiently. Domain-specific datasets are used to fine-tune a general model, resulting in multiple specialized TFMs customized for applications within specific subdomains. Data types, time granularity, and task similarities determine the categorization of these subdomains. This approach ensures that the specialized TFMs are well-suited for their respective applications' specific requirements and nuances. 
Figure \ref{fig:specialized_tfm} demonstrates fine-tuning a general TFM.

\begin{figure}[ht]
\centering
    \includegraphics[width =.9\linewidth]{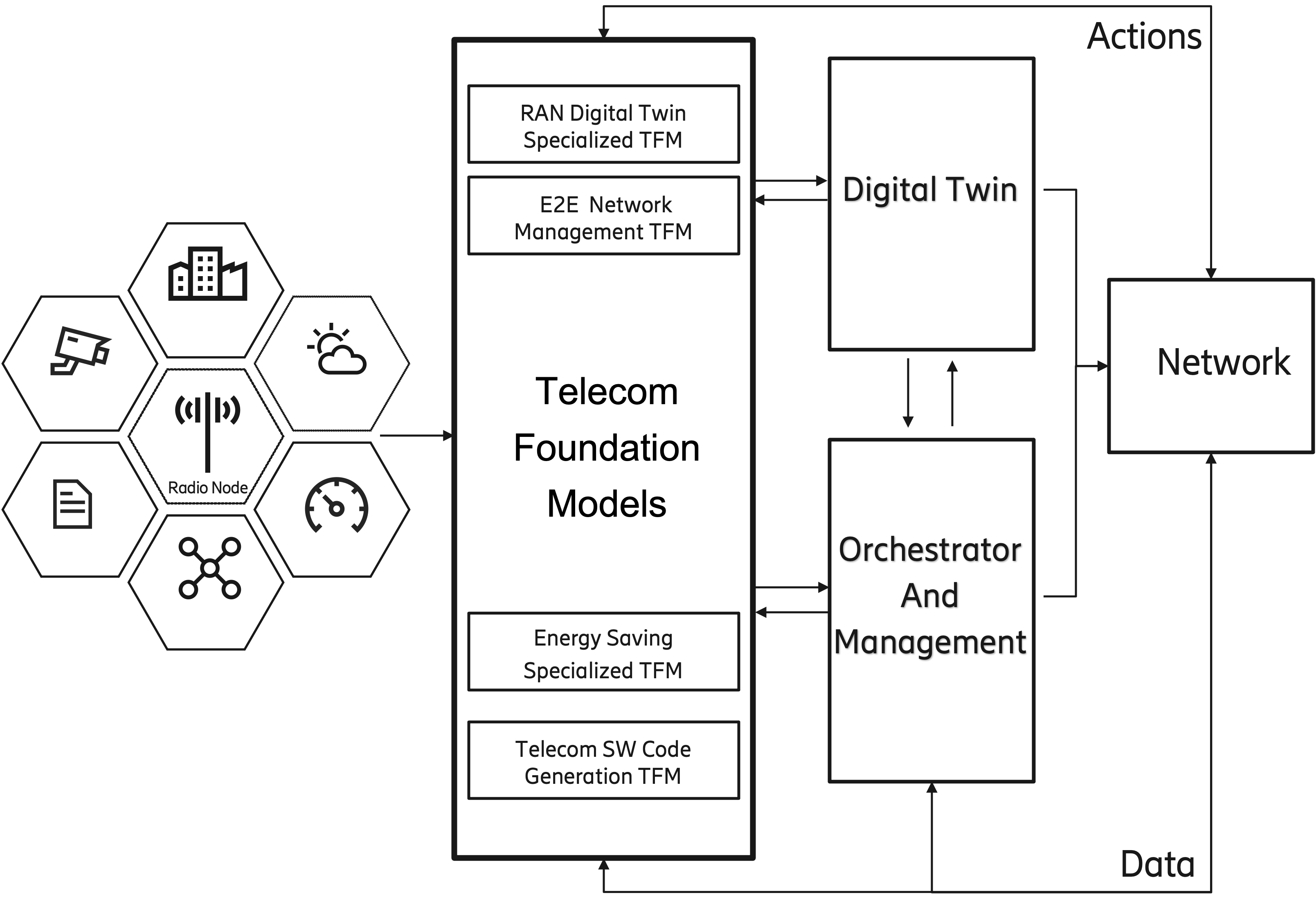}
    \caption{Specialized TFMs Architecture and Orchestration}
    \label{fig:tfm_orchestration}
\end{figure}

Specialized TFMs can be deployed in the central nodes or at the edge. Specialized TFMs can have different deployment requirements due to task description, HW, architectural limitations, required resources, latency requirement, etc. Due to impediments such as privacy, storage, and resource limitation, having one central model is always challenging. One possible solution to mitigate such challenges is to use recent distributed machine learning techniques such as \textit{Split Learning}. The overall architecture of TFM deployment, including data sources, application areas, and the orchestrator, is depicted in Figure~\ref{fig:tfm_orchestration}.  Since there are multiple specialized TFMs for different tasks and they might be dependent on each other outputs, there is a need to add \textit{Orchestration Layer} to the TFMs architecture design to ensure seamless network operation, provide communication flexibility, and mitigate conflicts. The orchestrator layer interconnects the TFMs and arranges the TFMs' tasks in a well-ordered, syncronized , and optimized pattern so that they function through accurate and automated repeated processes.

\section{TFM for Telecom Applications and Standardization Frameworks}
This section explores the potential of using TFMs in recent telecom standardization efforts.

\subsection{Intent-Based Networking}

The expanding scope of 5G and its diverse applications and requirements pose new challenges to telecom operations, especially in RAN. In this context, business intent and automated operations guarantee enterprise resilience. In telecom, \textit{"intent"} signifies the predefined objectives or desired behaviors expected from customers. An intent describes “what” needs to be achieved without identifying “how” to achieve them. Intent must be quantifiable from network data to measure and evaluate the fulfillment result. Intent-based networks revolutionize network operations by autonomously interpreting these business intents—such as improving network quality—into actions. By automating the process of integrating customer/provider/operator intent, translating it, and activating network features, these networks continuously align with business goals without human intervention. Using TFM, business intents are processed into service-level intents and translated into network KPIs. The FMs use these translated KPIs to propose and trigger actions in the network. These actions primarily affect the RAN, which bridges end users and business owners. Technological advancements in RAN to support diverse QoS requirements, such as high throughput and low latency, have increased the network's complexity. 

Any action in the network may have an instantaneous or long-term impact on the network performance. Therefore, there is a need to assess the action and avoid the risk of performance degradation. Digital Twin (DT) \cite{DT} emerges as a potential solution to bridge this gap. DT serves as a virtual representation of the physical telecom network that enables operators to experiment with new techniques and configurations without risking the physical network infrastructure. Creating DTs accurately models network functions and ensures efficient synchronization between physical and digital entities \cite{khan2022digital}. 

\begin{figure}[ht]
\centering
    \includegraphics[width =0.9\linewidth]{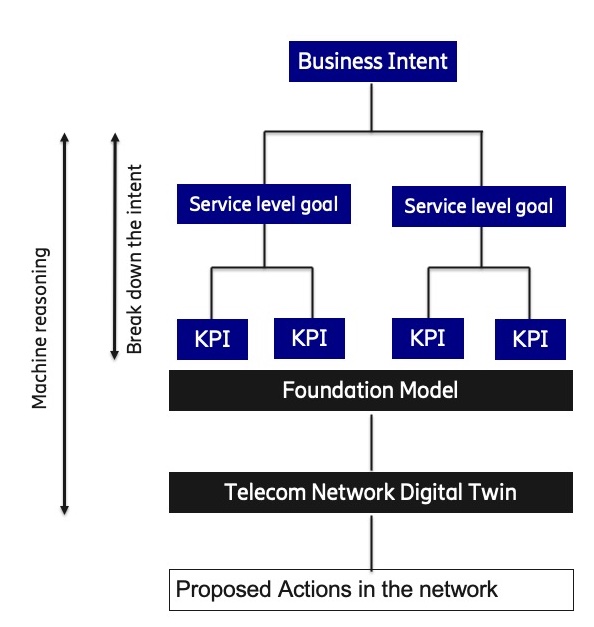}
    \caption{DT-assisted intent-based networking with TFMs}
    \label{fig:dt_vs_intent}
\end{figure}

The proposed actions by TFMs to adjust the network for satisfying intents can be assessed in DTs and then applied in the network. Moreover, TFMs can facilitate remote collaboration among distributed DTs and synchronize them with networks. Figure \ref{fig:dt_vs_intent} illustrates the interaction between DT and intent-based networking.

% % \subsubsection{Network Evolution and Deployment}

% %\textit{Self-Organizing Network}
% The continuous evolution of the network as well as technological advancements like O-RAN necessitates end-to-end network monitoring. Network monitoring allows for interaction between network segments and real-time assessment of metrics like latency and throughput. TFMs enhance this monitoring by proactively identifying congestion points and inefficient resource usage, enabling optimized resource allocation for improved network efficiency. This information can be used in network optimization applications such as slicing, energy saving, traffic steering, etc.

\subsection{Network Optimization}

The continuous evolution of the network, as well as technological advancements like O-RAN, necessitates end-to-end network monitoring. Network monitoring allows interaction between network segments and real-time assessment of intent metrics like latency and throughput. TFMs enhance this monitoring by proactively identifying congestion points and inefficient resource usage, enabling optimized resource allocation for improved network efficiency. This information can be used to demonstrate intent satisfaction and network optimization applications such as slicing, energy saving, traffic steering, etc.
Network optimization is a complex process that requires interactions among network segments. End-to-end optimization depends on models outlining interactions among multiple features. These models require skilled engineers possessing extensive domain expertise to identify pertinent aspects. Operators can benefit from using TFM-assisted DT to enhance this process and interact between network segments and parameter configurations \cite{ConfigurationRecGNN}.

\subsection{Network Slicing}

Network slicing empowers operators to optimize resource utilization and deliver diverse, scalable, differentiated services. A slice refers to dedicated end-to-end network resources.  Each slice represents a virtualized and independent logical network tailored to specific intents, operating on the same physical network infrastructure \cite{slicing}. 
According to Figure \ref{fig:dt_vs_intent}, the business intent is broken down into different intents. This splitting can be vertical in network architecture, meaning that each network segment has its intent to satisfy. However, the intent can be split into horizontal domains, meaning each service has its intent. Network slicing should consider both intents. Moreover, it needs to consider activated features in the network while facilitating slice resource allocation. In this context, TFM can collect the intents of all slices, activate network features along with network details, and recommend the necessary end-to-end resources for each slice. TFM passes the dedicated resources information to the DT to assess whether dedicated resources are insufficient, under-utilized, or adequate. If the dedicated resources are inadequate or under-utilized, TFM can submit a slice reconfiguration request. Consequently, TFMs offer slicing strategies that ensure intent feasibility and satisfaction per slice, prolonging the slice life cycle, enhancing resource utilization, and reducing energy consumption.

\subsection{Network Healing}

The complexities of deploying heterogeneous and dense telecom networks present a challenge to ensuring reliability, effectively finding optimal configuration parameters for resilient networks, and easily finding faults. TFMs promise to achieve self-healing networks capable of autonomously localizing faults, performing self-maintenance, and self-organizing without human intervention.

Troubleshooting efforts \cite{TroubleshootingGNN} can benefit from TFM-assisted DT to run tests \cite{TestCodeGenAI} in realistic setup. With the diverse deployment scenarios in Telecom networks and the large number of deployed products that exhibit a wide diversity, TFMs can help to overcome potential incidents, and autonomously perform remedial actions have the ability. The advntage of using TFMs originates from the ability to combine different data modalities to attain a holistic and context-aware understanding of network failures, enable better fault management, and achieve trustworthy and resilient networks. Moreover, troubleshooting efforts are often conducted in a distributed manner by multiple experts and multiple spoken languages to document issues around the globe. Language models have shown promising results \cite{bosch2022trdi} to remove language barriers and democratize expert knowledge through better knowledge search and troubleshooting recommendations from solved incidents.

\subsection{AI-powered Network APIs} 

Telecom infrastructure generates vast temporal and spatial digital traces through device interactions, offering a comprehensive insight into digital interactions. The design focuses on context awareness in telecom networks and aims to understand environmental, temporal, and situational factors impacting network behaviors. Leveraging this ubiquitous telecom data allows operators to explore new revenue streams. TFMs promise to unlock new ways to monetize Telecom Data through APIs. For instance, in smart city applications, historical real-time network performance data collected from transportation devices can provide invaluable insights into decision-making processes for autonomous vehicles. TFMs can utilize combined UE measurements and device characteristics to optimize route selection based on signal quality, further aiding in tasks like constructing maps, updating transportation routes, estimating traffic, and predicting travel times influenced by traffic conditions. Enabling other applications in other verticals (e.g.,  smart manufacturing, transportation sector, etc.) necessitates rethinking traditional telecom approaches, propelling us into a new era fueled by \textit{AI-as-services}.

% \textit{Zero-touch Networks}
% \textit{Trustworthy Networks}
% \textit{Resilient Networks}

\section{Future Trends and Open Issues}

Despite TFMs' potential to achieve AI-centric decision-making and network operations, several critical challenges remain unsolved. These challenges necessitate further investigation to realize the benefits of TFMs in the telecom industry fully.

\subsection{Scalability and Efficiency}
%\subsection{Edge Intelligence with TFMs}
Training FMs is a time-intensive process that demands significant computational resources, high power consumption, and specialized HW/SW infrastructure for model training. Moreover, deploying FMs can pose substantial computational demands, particularly regarding inference speed and model size. Consequently, the deployment of such models in real-time scenarios is still limited. Solutions such as model compression techniques (e.g., pruning and quantization) and distributed machine learning techniques (e.g., split learning, federated learning) can address these challenges.  Model compression can lead to a trade-off between speed and accuracy, and distributed techniques can introduce communication overhead and synchronization issues. Therefore, ongoing research and development are crucial to enhance these methods and develop new approaches to make FMs more efficient and practical for real-time deployment in telecom networks. Despite these strategies, achieving high-speed inference with comparable performance remains a significant challenge.

\subsection{Transparency and Interpretability}
The complexity and scale of FMs present a significant challenge when applying FMs to the telecom domain. FMs still struggle with randomness when generating output. Hence, the trustworthiness and explainability of TMFs remain a challenge that hinders adopting the long-term vision of TFMs-native framework for telecom and fostering trust in the operation. It is aligned with current standarization. For instance, 3GPP has an ambition to create AI/ML processes with full interoperability among all the components of a 5G RAN system.

%Moreover, when TFMs undergo training with new data, they tend to overlook information associated with previously encountered data to assimilate new information. This occurrence, known as catastrophic forgetting, poses a significant challenge in sequential and lifelong learning scenarios, such as acquiring proficiency in new tasks and domains.
%\cite{zhou2023comprehensive} \cite{bosch2022trdi}

% \subsection{RAN Multi-Tasking Operations} 
% With several uncertainties surrounding deployment environments, many applications in RAN require a multi-modal, multi-step, multi-task, and collaborative decision-making process. Satisfying intent in such tasks remains a significant challenge. Reinforcement Learning  Foundation Models RL FMs are emerging techniques that can address these challenges. However, RL FMs are still in the early stages and face challenges, including data heterogeneity and diverse data modalities. Investigating these approaches and preparing them for telecom use cases is necessary. 

% Some telecom applications, such as dynamic spectrum resource sharing, require control loops with low latency inference time. This can be challenging for specialized TFMs and may limit their applicability. Additionally, ensuring reliability and robustness in decision-making processes under varying and unpredictable conditions adds another layer of complexity. To overcome these obstacles, ongoing research and development are essential to refine RL FMs, enhance their performance, and ensure they can meet the stringent demands of real-time RAN operations. 

\subsection{Time-critical Applications} 
Reducing latency in TFMs is essential for real-time applications such as dynamic spectrum resource sharing, resource allocation, user associations, handover management, and carrier aggregation. These use cases require control loops with low latency inference time. This can be challenging for specialized TFMs and may limit their applicability. Hence, latency must be minimized to ensure efficient and timely actions. A key strategy is to keep processing close to the data and TFM deployment location, reducing the need for extensive data transfers between compute resources and storage. This approach enhances performance, lowers costs, and mitigates data security risks, making it critical for effectively deploying TFMs in RAN.
%Reducing latency in TFMs is essential for real-time applications such as resource allocation, user associations, handover management, and carrier aggregation. Latency, the time it takes for the model to make predictions after receiving input data, must be minimized to ensure efficient and timely actions. A key strategy is to keep processing close to the data and TFM deployment location, reducing the need for extensive data transfers between compute resources and storage. This approach enhances performance, lowers costs, and mitigates data security risks, making it critical for effectively deploying TFMs in RAN.

%\subsection{Product Efficiency Improvement}

%The advent of network technologies and the introduction of Open RAN present a challenge for telecom vendors and mobile operators in effectively developing resilient SW, improving the manufacturing process, and enhancing testing. In recent years, FMs have excelled in generating code \cite{TestCodeGenAI}. Capitalizing on contemporary advancements in FMs for code, the telecom industry can enhance development processes that require generating documentation,  test cases,  synthetic data for tests, error handling, commit validation, and code translation, among other tasks.  TFMs show immense promise for development and are envisaged to substantially participate in advancing smart manufacturing and enhancing processes to improve quality, efficiency, cost, and agility for HW manufacturing.

\section{Conclusion}
This article explored the untapped potential of how TFMs can shape the future of mobile networks. In particular, we explored how TFMs can be effectively used to develop, upgrade, operate, and manage networks. Moreover, we identified key ingredients to design and train FMs given the wide telecom ecosystem to rethink our traditional engineering approaches. 

TFMs present an immense opportunity to significantly shape telecom networks on their journey towards 6G and beyond, simplify network operations, enhance development productivity, promote sustainability, expand the value chain, and ultimately boost business profitability. 

Through our discussions, it has become apparent that TFMs face challenges in terms of scalability in resource-constrained environments. Despite these hurdles, TFMs offer to build efficient, resilient, and trustworthy networks.

% I don't know if we should add this phrase
%The paper has aimed to provide an overview of the untapped potential of foundation models for telecom applications and motivate the need for an aligned view and shared practices to leverage telecom Foundation Models.

%\printbibliography

\bibliographystyle{IEEEtran}
\bibliography{bibliography}
\end{document}